\documentclass[12pt,showpacs,showkeys,amsmath,amssymb]{revtex4}
\usepackage[T1]{fontenc}
\usepackage[latin1]{inputenc}
\usepackage{amssymb}

\makeatletter

\newcommand{\diag}{\operatorname{diag}}

\makeatother
\begin{document}

\title{Toward a Nonlocal Theory of Gravitation}

\date{\today}

\author{Bahram Mashhoon}

\affiliation{Department of Physics and Astronomy, University of Missouri-Columbia,
Columbia, Missouri 65211, USA\\
mashhoonb@missouri.edu}

\begin{abstract}
The nonlocal theory of accelerated systems is extended to linear gravitational
waves as measured by accelerated observers in Minkowski spacetime.
The implications of this approach are discussed. In particular, the
nonlocal modifications of helicity-rotation coupling are pointed out
and a nonlocal wave equation is presented for a special class of uniformly
rotating observers. The results of this study, via Einstein's heuristic
principle of equivalence, provide the incentive for a nonlocal classical
theory of the gravitational field. 
\end{abstract}

\pacs{04.20.Cv, 11.10.Lm}

\keywords{Gravitation, nonlocality.}

\maketitle

\section{Introduction }

In a recent paper~\cite{key-1}, the spin-rotation-gravity coupling
has been worked out in detail for linear gravitational waves. In particular,
it has been demonstrated that a plane monochromatic linear gravitational
wave of frequency $\omega$ propagating in Minkowski spacetime has
frequency $\omega'$,\begin{equation}
\omega'=\gamma(\omega\mp2\Omega),\label{eq:1}\end{equation}
 as measured by an observer rotating uniformly with frequency $\Omega$
about the direction of propagation of the incident radiation. Here
$\gamma$ is the Lorentz factor of the observer and the upper (lower)
sign corresponds to positive (negative) helicity radiation. More generally,
\begin{equation}
\omega'=\gamma(\omega-m\Omega),\label{eq:2}\end{equation}
 where $m=0,\pm1,\pm2,\dots,$ is the total (orbital plus spin) angular
momentum parameter in the case of oblique incidence~\cite{key-1}.

Equations~\eqref{eq:1} and \eqref{eq:2} are the spin-2 analogues
of similar results for electromagnetic radiation that have been discussed
in detail in~\cite{key-2}. In Eq.~\eqref{eq:2}, $\omega'$ can
be zero or negative. A negative $\omega'$ cannot be excluded due
to the absolute character of the observer's rotation. However, in
the case of $\omega'=0$, there is no experimental evidence to suggest
that a basic radiation field could ever stand completely still with
respect to any observer. In the derivation of Eqs.~\eqref{eq:1}
and \eqref{eq:2}, the standard theory of relativity based on the
hypothesis of locality has been employed. A consequence of this assumption
is that the gravitational wave could stand completely still for $\omega=m\Omega$
in Eq.~\eqref{eq:2}. For instance, by a mere rotation of frequency
$\omega/2$ in the positive sense about the direction of propagation
of a normally incident positive helicity gravitational wave, the field
becomes completely static in accordance with Eq.~\eqref{eq:1}. Under
Lorentz transformations, however, a linear gravitational radiation
field can never stand still with respect to an \textit{inertial} observer;
indeed, this is the case for all basic radiation fields. Generalizing
this circumstance to all observers, a nonlocal theory of accelerated
observers has been developed that goes beyond the hypothesis of locality
and is consistent with all observational data available at present~\cite{key-2,key-3}.
Specifically, the nonlocal electrodynamics of rotating systems has
been successfully tested indirectly via the agreement of the nonlocal
theory's predictions with standard quantum mechanical results for
the electromagnetic interactions of rotating electrons in the correspondence
limit {[}2]. Moreover, the postulates of the nonlocal theory forbid
the existence of a fundamental scalar (or pseudoscalar) radiation
field in nature in agreement with observation.

Let $\psi(x)$ be a basic radiation field in Minkowski spacetime and
imagine an accelerated observer that measures this field as a function
of proper time $\tau$ along its worldline. Let $\hat{\Psi}(\tau)$
be the result of such a measurement. According to the hypothesis of
locality, the accelerated observer is pointwise equivalent to an otherwise
identical hypothetical momentarily comoving inertial observer. Let
$\hat{\psi}(\tau),$\begin{equation}
\hat{\psi}(\tau)=\Lambda(\tau)\psi(\tau),\label{eq:3}\end{equation}
 where $\Lambda$ is a matrix representation of the Lorentz group,
be the measured field according to the infinite set of such momentarily
comoving inertial observers; therefore, the hypothesis of locality
would require that $\hat{\Psi}(\tau)=\hat{\psi}(\tau)$. On the other
hand, the most general linear relationship between $\hat{\Psi}$ and
$\hat{\psi}$ that is consistent with causality can be expressed as\begin{equation}
\hat{\Psi}(\tau)=\hat{\psi}(\tau)+\int_{\tau_{0}}^{\tau}K(\tau,\tau')\hat{\psi}(\tau')d\tau'\label{eq:4}\end{equation}
 for $\tau\geq\tau_{0}$, where $\tau_{0}$ is the instant at which
the acceleration is turned on. Equation~\eqref{eq:4}, which expresses
the nonlocality of field determination by an accelerated observer,
involves a weighted average over the past worldline of the observer
and is thus compatible with the ideas put forth by Bohr and Rosenfeld~\cite{key-4}.
The kernel $K$ must clearly vanish in the absence of acceleration.
It follows from the results of Volterra~\cite{key-5} and Tricomi~\cite{key-6}
that the relationship between $\psi(\tau)$ and $\hat{\Psi}(\tau)$
is unique in the space of functions of physical interest.

To ensure that a basic radiation field never stands completely still
with respect to an accelerated observer, i.e. $\hat{\Psi}$ is variable
when $\psi$ is, we associate a constant $\hat{\Psi}$ with a constant
$\psi$. The Volterra-Tricomi uniqueness theorem~\cite{key-5,key-6}
then excludes the possibility that a constant $\hat{\Psi}$ could
ever result from a variable $\psi$. Our postulate thus implies the
following integral equation for the kernel $K$ \begin{equation}
\Lambda(\tau_{0})=\Lambda(\tau)+\int_{\tau_{0}}^{\tau}K(\tau,\tau')\Lambda(\tau')d\tau'.\label{eq:5}\end{equation}
 The solutions of this equation have been investigated in detail~\cite{key-7,key-8}.
It turns out that the only acceptable solution is given by $K(\tau,\tau')=k(\tau')$,\begin{equation}
k(\tau)=-\frac{d\Lambda(\tau)}{d\tau}\Lambda^{-1}(\tau).\label{eq:6}\end{equation}
 The consequences of this theory for nonlocal electrodynamics have
been worked out in detail~\cite{key-2,key-3}. Moreover, $k=0$ for
a scalar (or pseudoscalar) field, which is therefore always local
and hence subject to the difficulty involving scalar (or pseudoscalar)
radiation standing completely still with respect to certain rotating
observers. It would be interesting to extend the nonlocal ansatz to
linear gravitational waves in Minkowski spacetime and explore some
of the consequences of the resulting theory. This is done in the rest
of this paper.

The spacetime metric associated with a linear gravitational wave is
given by $g_{\mu\nu}=\eta_{\mu\nu}+h_{\mu\nu}(x^{\alpha})$, where
$(\eta_{\mu\nu})=\diag(-1,1,1,1)$ is the Minkowski metric tensor,
$x^{\alpha}=(ct,x,y,z)$ and $h_{\mu\nu}$ is a sufficiently small
perturbation subject to the gauge transformation\begin{equation}
h_{\mu\nu}\mapsto h_{\mu\nu}+\epsilon_{\mu,\nu}+\epsilon_{\nu,\mu}\label{eq:7}\end{equation}
 due to an infinitesimal transformation of inertial coordinates $x^{\mu}\mapsto x^{\mu}-\epsilon^{\mu}$.
The gauge-invariant field strength is given by the Riemann curvature
tensor\begin{equation}
R_{\mu\nu\rho\sigma}=\frac{1}{2}(h_{\mu\sigma,\nu\rho}+h_{\nu\rho,\mu\sigma}-h_{\nu\sigma,\mu\rho}-h_{\mu\rho,\nu\sigma}).\label{eq:8}\end{equation}
 In what follows, we will regard $h_{\mu\nu}$ and $R_{\mu\nu\rho\sigma}$
as fields defined in a global inertial frame in Minkowski spacetime.
It proves useful to introduce the trace-reversed wave amplitude $\bar{h}_{\mu\nu}=h_{\mu\nu}-\frac{1}{2}\eta_{\mu\nu}h$,
where $h=\eta^{\mu\nu}h_{\mu\nu}$. Imposing the transverse gauge
condition $\bar{h}_{\;\;\;\;,\nu}^{\mu\nu}=0$, the source-free gravitational
field equation reduces to the wave equation $\square\bar{h}_{\mu\nu}=0$
in this case~\cite{key-9}. The remaining gauge freedom is usually
restricted by introducing the transverse-traceless gauge in which
the conditions $h=0$ and $h_{0\mu}=0$ are further imposed.

It is well known that the treatment of gravitational waves outlined
above, namely, the linear approximation of general relativity for
free gravitational fields on a Minkowski spacetime background admits
of an alternative interpretation: It can be regarded as a Lorentz-invariant
theory of a free linear massless spin-2 field in special relativity.
This latter approach --- to which the nonlocal theory of accelerated
systems~\cite{key-2,key-3} is directly applicable --- is adopted
in the rest of this paper.

It is important to recognize that the nonlocal ansatz can be applied
either to the gravitational field ($R_{\mu\nu\rho\sigma}$) or the
gravitational wave potential ($h_{\mu\nu}$ or $\bar{h}_{\mu\nu}$)
resulting in two distinct but closely related approaches. The situation
here is completely analogous to the electromagnetic case~\cite{key-10}.
For the sake of simplicity, we choose the latter alternative in what
follows. The potential as measured by an arbitrary accelerated observer
in Minkowski spacetime is given by\begin{equation}
h_{\hat{\alpha}\hat{\beta}}=h_{\mu\nu}\lambda_{\;\;\hat{\alpha}}^{\mu}\lambda_{\;\;\hat{\beta}}^{\nu},\label{eq:9}\end{equation}
 where $\lambda_{\;\;\hat{\alpha}}^{\mu}$ is the orthonormal tetrad
associated with the observer. Our nonlocal ansatz~\eqref{eq:4} for
the $h_{\mu\nu}$ then takes the Lorentz-invariant form\begin{equation}
H_{\hat{\alpha}\hat{\beta}}(\tau)=h_{\hat{\alpha}\hat{\beta}}(\tau)+\int_{\tau_{0}}^{\tau}k_{\hat{\alpha}\hat{\beta}}^{\;\;\;\;\hat{\gamma}\hat{\delta}}(\tau')h_{\hat{\gamma}\hat{\delta}}(\tau')d\tau',\label{eq:10}\end{equation}
 where $H_{\hat{\alpha}\hat{\beta}}$ is the gravitational wave amplitude
as measured by the accelerated observer.

In general, the symmetric tensor $h_{\mu\nu}$ (or $\bar{h}_{\mu\nu}$)
has ten independent components. We arrange these in a column vector
$\psi$ such that Eq.~\eqref{eq:9}, or the analogous one for $\bar{h}_{\mu\nu}$,
can take the form of Eq.~\eqref{eq:3} with a $10\times10$ matrix
$\Lambda$. Specifically, $\hat{\psi}_{A}=\Lambda_{A}^{\;\; B}\psi_{B}$,
where indices $A$ and $B$ belong to the set $\left\{ 00,01,02,03,11,12,13,22,23,33\right\} $.
For the sake of definiteness, we will henceforth assume that $\psi$
represents $h_{\mu\nu}$.

It is worthwhile to work out explicitly the nonlocal theory of linear
gravitational waves for certain accelerated observers. Section~\ref{sec:2}
is devoted to uniformly rotating observers; then, the nonlocal results
are employed in Section~\ref{sec:3} to re-examine the status of
helicity-rotation coupling for gravitational radiation. For a special
class of uniformly rotating observers, the nonlocal gravitational
wave equation is derived in Section~\ref{sec:3a}. Section~\ref{sec:4}
explores the case of translationally accelerated observers. The implications
of the nonlocal treatment of linear gravitational waves via Einstein's
principle of equivalence are discussed in Section~\ref{sec:5}. Some
of the computational details are relegated to the appendices.

\section{Uniformly rotating observer\label{sec:2}}

Consider an observer that for $t<0$ moves uniformly in the $(x,y)$
plane of an inertial system of coordinates such that $x=r$ and $y=r\Omega t$,
where $r$ and $\Omega$ are positive constants. Suppose that at $t=0$
the observer begins to move on a circle of radius $r$ with $x=r\cos\phi$,
$y=r\sin\phi$ and $z=0$. Here $\phi=\Omega t=\gamma\Omega\tau$,
where $\gamma$ is the observer's Lorentz factor that corresponds
to $\beta=r\Omega/c$. For $\tau>0$, the observer's orthonormal tetrad
frame~\cite{key-1,key-2} is given by\begin{align}
\lambda_{\;\;\hat{0}}^{\mu} & =\gamma(1,-\beta\sin\phi,\beta\cos\phi,0),\label{eq:11}\\
\lambda_{\;\;\hat{1}}^{\mu} & =(0,\cos\phi,\sin\phi,0),\label{eq:12}\\
\lambda_{\;\;\hat{2}}^{\mu} & =\gamma(\beta,-\sin\phi,\cos\phi,0),\label{eq:13}\\
\lambda_{\;\;\hat{3}}^{\mu} & =(0,0,0,1).\label{eq:14}\end{align}
 In this case, Eq.~\eqref{eq:9} is written out in detail in Appendix~\ref{sec:A}
and thereby $\Lambda$ can be immediately constructed. The kernel
$k$ can then be determined using Eq.~\eqref{eq:6}. The general
case will not be treated here; instead, we focus attention on the
simple case of the observer that is at rest at the spatial origin
of coordinates and refers its measurements to uniformly rotating axes,
i.e. $r=0$, so that $\beta=0$ and $\gamma=1$ in Eqs.~\eqref{eq:11}
and \eqref{eq:13}. In this case $\Lambda$ has a block diagonal form,
$\Lambda=\diag(1,R,1,T,1)$, where $R$ is the $2\times2$ rotation
matrix\begin{equation}
R(\phi)=\left[\begin{array}{cc}
\cos\phi & \sin\phi\\
-\sin\phi & \cos\phi\end{array}\right]\label{eq:15}\end{equation}
 and $T$ is the $5\times5$ matrix\begin{equation}
T(\phi)=\left[\begin{array}{ccccc}
\cos^{2}\phi & \sin2\phi & 0 & \sin^{2}\phi & 0\\
-\frac{1}{2}\sin2\phi & \cos2\phi & 0 & \frac{1}{2}\sin2\phi & 0\\
0 & 0 & \cos\phi & 0 & \sin\phi\\
\sin^{2}\phi & -\sin2\phi & 0 & \cos^{2}\phi & 0\\
0 & 0 & -\sin\phi & 0 & \cos\phi\end{array}\right].\label{eq:16}\end{equation}
 We note that $\det R=\det T=1$ and\begin{equation}
R^{-1}(\phi)=R(-\phi),\quad T^{-1}(\phi)=T(-\phi).\label{eq:17}\end{equation}

The kernel $k$ can be easily determined in this case, since $\Lambda^{-1}=\diag(1,R(-\phi),1,T(-\phi),1)$.
It then turns out that $k$ is a constant matrix $\left(k_{m,n}\right)$,
$m,n=1,\dots,10$, with nonzero elements given by\begin{align}
k_{2,3} & =k_{6,8}=k_{7,9}=-\Omega,\label{eq:18}\\
k_{3,2} & =k_{6,5}=k_{9,7}=\Omega,\label{eq:19}\\
-k_{5,6} & =k_{8,6}=2\Omega.\label{eq:20}\end{align}

The kernel can be used to determine the nonlocal relationship, based
on the ansatz~\eqref{eq:10}, between the measured components of
the gravitational potential, denoted by $H_{\hat{\alpha}\hat{\beta}}$,
and the components $h_{\hat{\alpha}\hat{\beta}}$ that are obtained
from the hypothesis of locality. The end result can be expressed as

\begin{equation}
H_{\hat{0}\hat{0}}=h_{\hat{0}\hat{0}},\quad H_{\hat{0}\hat{3}}=h_{\hat{0}\hat{3}},\quad H_{\hat{3}\hat{3}}=h_{\hat{3}\hat{3}},\label{eq:21}\end{equation}
 \begin{subequations}\begin{align}
H_{\hat{0}\hat{1}}+H_{\hat{0}\hat{2}} & =h_{\hat{0}\hat{1}}+h_{\hat{0}\hat{2}}+\Omega\int_{0}^{\tau}(h_{\hat{0}\hat{1}}-h_{\hat{0}\hat{2}})d\tau',\label{eq:22a}\\
H_{\hat{0}\hat{1}}-H_{\hat{0}\hat{2}} & =h_{\hat{0}\hat{1}}-h_{\hat{0}\hat{2}}-\Omega\int_{0}^{\tau}(h_{\hat{0}\hat{1}}+h_{\hat{0}\hat{2}})d\tau',\label{eq:22b}\end{align}
 \end{subequations}\begin{subequations}\begin{align}
H_{\hat{1}\hat{1}}+H_{\hat{2}\hat{2}} & =h_{\hat{1}\hat{1}}+h_{\hat{2}\hat{2}},\label{eq:23a}\\
H_{\hat{1}\hat{2}} & =h_{\hat{1}\hat{2}}+\Omega\int_{0}^{\tau}(h_{\hat{1}\hat{1}}-h_{\hat{2}\hat{2}})d\tau',\label{eq:23b}\\
H_{\hat{1}\hat{1}}-H_{\hat{2}\hat{2}} & =h_{\hat{1}\hat{1}}-h_{\hat{2}\hat{2}}-4\Omega\int_{0}^{\tau}h_{\hat{1}\hat{2}}\; d\tau',\label{eq:23c}\end{align}
 \end{subequations}\begin{subequations}\begin{align}
H_{\hat{1}\hat{3}}+H_{\hat{2}\hat{3}} & =h_{\hat{1}\hat{3}}+h_{\hat{2}\hat{3}}+\Omega\int_{0}^{\tau}(h_{\hat{1}\hat{3}}-h_{\hat{2}\hat{3}})d\tau',\label{eq:24a}\\
H_{\hat{1}\hat{3}}-H_{\hat{2}\hat{3}} & =h_{\hat{1}\hat{3}}-h_{\hat{2}\hat{3}}-\Omega\int_{0}^{\tau}(h_{\hat{1}\hat{3}}+h_{\hat{2}\hat{3}})d\tau'.\label{eq:24b}\end{align}
 \end{subequations} These equations, combined with the results of
Appendix~\ref{sec:A}, can be employed to determine the nonlocal
modifications of the helicity-rotation coupling for gravitational
waves incident on the special rotating observer that occupies the
origin of spatial coordinates.

\section{Helicity-rotation coupling\label{sec:3}}

Consider the reception of a plane monochromatic gravitational wave
of definite helicity incident along the $z$ axis by the rotating
observer that is fixed at the origin of spatial coordinates. In the
transverse-traceless gauge, the wave amplitude is given by the real
part of\begin{equation}
(h_{ij})=A(e_{\oplus}\pm ie_{\otimes})e^{i\omega(-t+z/c)}.\label{eq:25}\end{equation}
 Here $A$ is a constant amplitude, the upper (lower) sign corresponds
to positive (negative) helicity radiation and the two independent
linear polarization states~\cite{key-1} are given by \begin{equation}
e_{\oplus}=\left[\begin{array}{ccc}
1 & 0 & 0\\
0 & -1 & 0\\
0 & 0 & 0\end{array}\right],\quad e_{\otimes}=\left[\begin{array}{ccc}
0 & 1 & 0\\
1 & 0 & 0\\
0 & 0 & 0\end{array}\right].\label{eq:26}\end{equation}
 For wave functions, the complex representation is employed throughout
as all operations involving gravitational waves are linear. Thus only
the real parts of the relevant quantities are of physical interest.

It follows from a detailed examination of the results of the previous
section that in this case\begin{align}
h_{\hat{\alpha}\hat{\beta}} & =e^{\pm2i\phi}h_{\alpha\beta}\label{eq:27}\\
\intertext{and}H_{\hat{\alpha}\hat{\beta}} & =F_{\pm}(\tau)h_{\hat{\alpha}\hat{\beta}},\label{eq:28}\\
\intertext{where}F_{\pm}(\tau) & =\frac{\omega\mp2\Omega e^{i\omega'\tau}}{\omega\mp2\Omega},\label{eq:29}\end{align}
 $\omega'=\omega\mp2\Omega$, $\tau=t$ and $z=0$. Equations~\eqref{eq:27}-\eqref{eq:29}
are simply the spin-2 analogues of the corresponding results that
have been discussed in detail in nonlocal electrodynamics---cf. Eqs.~\eqref{eq:12}-\eqref{eq:17}
of~\cite{key-2}. Specifically, for the case of resonance involving
an incident positive helicity wave of frequency $\omega\mapsto2\Omega$,
we find that as $\omega'\mapsto0$, $F_{+}\mapsto f_{+}=1-2i\Omega\tau$;
this linear divergence with time can be avoided with a finite incident
wave packet. On the other hand, for an incident negative helicity
wave of $\omega=2\Omega$, $\omega'=4\Omega$ and $F_{-}$ becomes
$f_{-}=\cos(2\Omega\tau)\exp(2i\Omega\tau)$.

Another direct consequence of nonlocality, evident in the factor $F_{\pm}$,
is that the amplitude of a positive helicity gravitational wave of
$\omega>2\Omega$ as measured by the rotating observer is enhanced
by a factor of $\omega/(\omega-2\Omega)$, while that of a negative
helicity wave is diminished by a factor of $\omega/(\omega+2\Omega)$.

The nonlocal aspects of linear gravitation developed here may be extended
to an inertial observer in the gravitomagnetic field of a rotating
mass via the gravitational Larmor theorem~\cite{key-1}. Moreover,
it should be remarked that for Earth-based gravitational-wave antennas
the effective rotation frequency is about $10^{-5}$Hz. The nonlocal
effects discussed here would then be ordinarily very small for incident
high-frequency gravitational waves with $\Omega/\omega<<1$; in fact,
nonlocality could only become significant near resonance.

The results that have been obtained thus far for the rotating observer
fixed at the origin of spatial coordinates may be simply extended
to a whole class of such observers that are fixed in space and differ
from each other only through their spatial positions. It turns out
to be simpler to deal with this class of uniformly rotating observers
than the class of observers whose tetrads are given by Eqs. (11)-(14).
In the following section, we present the nonlocal gravitational field
equation for the class of spatially fixed rotating observers.

\section{Nonlocal wave equation\label{sec:3a}}

Imagine observers that are always at rest in a global inertial frame
and refer their measurements to the standard inertial axes for $-\infty<t<0$;
for $t\geq0$, however, they employ axes that rotate uniformly about
the $z$ axis with frequency $\Omega$. Thus for $t\geq0$, each such
observer carries a tetrad frame given by Eqs.~\eqref{eq:11}-\eqref{eq:14}
with $\beta=0$ and $\gamma=1$. The purpose of this section is to
develop the Lorentz-invariant nonlocal gravitational wave equation
for this special class of noninertial observers.

It is a general consequence of Eq.~\eqref{eq:10} that for $\tau>\tau_{0}$,\begin{equation}
h_{\alpha\beta}(\tau)=H_{\alpha\beta}(\tau)+\int_{\tau_{0}}^{\tau}\tilde{r}_{\alpha\beta}^{\;\;\;\;\gamma\delta}(\tau,\tau')H_{\gamma\delta}(\tau')d\tau',\label{eq:30a}\end{equation}
where $\tilde{r}$ is a variant of the resolvent kernel and has been
discussed in detail in~\cite{key-10}. It has been shown in Appendix~C
of \cite{key-10} that if $k$ is a constant kernel, then $\tilde{r}$
is constant as well and is given by\begin{equation}
\tilde{r}=-\Lambda^{-1}(\tau_{0})k\Lambda(\tau_{0}).\label{eq:31b}\end{equation}

For the special class of rotating observers under consideration here,
$\tau=t$, $\tau_{0}=0$, $k$ is a constant matrix and its nonzero
elements are given in Eqs.~\eqref{eq:18}-\eqref{eq:20}. Moreover,
it is clear from Eqs.~\eqref{eq:15}-\eqref{eq:16} that $\Lambda(0)$
is the identity matrix; hence, it follows from Eq.~\eqref{eq:31b}
that in this case\begin{equation}
\tilde{r}=-k.\label{eq:32a}\end{equation}
Thus the explicit form of the ten independent equations contained
in Eq.~\eqref{eq:30a} may be obtained from Eqs.~\eqref{eq:21}-\eqref{eq:24b}
by making the formal replacement $(H_{\hat{\alpha}\hat{\beta}},h_{\hat{\alpha}\hat{\beta}},\Omega)\mapsto(h_{\alpha\beta},H_{\alpha\beta},-\Omega)$.
It is interesting to note that the form of Eqs.~\eqref{eq:21}-\eqref{eq:24b}
remains the same if all of the indices are raised; the same is true
for the explicit form of Eq.~\eqref{eq:30a} in the case under consideration
here.

To express Eq.~\eqref{eq:30a} for the \textit{special class of rotating
observers}, it proves convenient to write\begin{equation}
h^{\alpha\beta}(t,\mathbf{x})=H^{\alpha\beta}(t,\mathbf{x})+\tilde{r}_{\;\;\;\;\gamma\delta}^{\alpha\beta}\int_{0}^{t}H^{\gamma\delta}(t',\mathbf{x})dt'\label{eq:33a}\end{equation}
for $t>0$, since the observers are fixed in space. Here the components
of $\tilde{r}$ are all constants proportional to $\Omega$. The substitution
of $h^{\alpha\beta}(t,\mathbf{x})$ in the equations that it satisfies
would then result, via Eq.~\eqref{eq:33a}, in the corresponding
equations for the nonlocal wave amplitude $H^{\alpha\beta}(t,\mathbf{x}).$

The wave function $h^{\alpha\beta}(t,\mathbf{x})$ is subject to the
gauge condition\begin{equation}
\left(h^{\alpha\beta}-\frac{1}{2}\eta^{\alpha\beta}h\right)_{\!,\beta}=0\label{eq:34}\end{equation}
and satisfies the wave equation\begin{equation}
\square h^{\alpha\beta}=0,\label{eq:35}\end{equation}
which follows from $\square\bar{h}_{\alpha\beta}=0$. Thus for $t>0$,
$H^{\alpha\beta}$ is subject to the gauge condition\begin{equation}
\left(H^{\alpha\beta}+\tilde{r}_{\;\;\;\;\gamma\delta}^{\alpha\beta}\int_{0}^{t}H^{\gamma\delta}(t',\mathbf{x})dt'\right)_{\!,\beta}=\frac{1}{2}\eta^{\alpha\beta}H_{,\beta},\label{eq:36}\end{equation}
where $H=\eta_{\alpha\beta}H^{\alpha\beta}$ and it turns out that
$H=h$ in this case. Moreover, $H^{\alpha\beta}$satisfies the nonlocal
wave equation\begin{equation}
\square H^{\alpha\beta}=\tilde{r}_{\;\;\;\;\gamma\delta}^{\alpha\beta}\left(\frac{\partial}{\partial t}H^{\gamma\delta}-\Delta\int_{0}^{t}H^{\gamma\delta}(t',\mathbf{x})dt'\right)\label{eq:37a}\end{equation}
for $t>0$.

The measurements of a class of linearly accelerated observers are
considered in the next section. In particular, we will rule out the
possibility of existence of a direct nonlocal coupling between the
helicity of gravitational radiation and linear acceleration.

\section{Linearly accelerated observers\label{sec:4}}

Consider the class of observers at rest in the background global inertial
system for $-\infty<t<0$. At $t=0$, the observers accelerate from
rest with acceleration $g(\tau)>0$ along the $z$ axis. Here $\tau$
is the proper time and $\tau=0$ at $t=0$. For $\tau\geq0$, the
orthonormal tetrad frame of the observers is given by\begin{align}
\lambda_{\;\;\hat{0}}^{\mu} & =(C,0,0,S), & \quad\lambda_{\;\;\hat{1}}^{\mu} & =(0,1,0,0),\label{eq:30}\\
\lambda_{\;\;\hat{2}}^{\mu} & =(0,0,1,0), & \quad\lambda_{\;\;\hat{3}}^{\mu} & =(S,0,0,C),\label{eq:31}\end{align}
 where $C=\cosh\theta$, $S=\sinh\theta$ and\begin{equation}
\theta=\frac{1}{c}\int_{0}^{\tau}g(\tau')d\tau'.\label{eq:32}\end{equation}

It follows from Eq.~\eqref{eq:9} that the wave amplitude, as measured
by the momentarily comoving observers, is given by\begin{equation}
h_{\hat{1}\hat{1}}=h_{11},\quad h_{\hat{1}\hat{2}}=h_{12},\quad h_{\hat{2}\hat{2}}=h_{22},\label{eq:33}\end{equation}
 \begin{subequations}\begin{align}
h_{\hat{0}\hat{0}}-h_{\hat{3}\hat{3}} & =h_{00}-h_{33},\label{eq:34a}\\
\frac{1}{2}(h_{\hat{0}\hat{0}}+h_{\hat{3}\hat{3}}) & =\frac{1}{2}(h_{00}+h_{33})\cosh2\theta+h_{03}\sinh2\theta,\label{eq:34b}\\
h_{\hat{0}\hat{3}} & =h_{03}\cosh2\theta+\frac{1}{2}(h_{00}+h_{33})\sinh2\theta,\label{eq:34c}\end{align}
 \end{subequations}\begin{subequations}\begin{align}
h_{\hat{0}\hat{1}}-h_{\hat{1}\hat{3}} & =(h_{01}-h_{13})e^{-\theta},\label{eq:35a}\\
h_{\hat{0}\hat{1}}+h_{\hat{1}\hat{3}} & =(h_{01}+h_{13})e^{\theta},\label{eq:35b}\end{align}
 \end{subequations}\begin{subequations}\begin{align}
h_{\hat{0}\hat{2}}-h_{\hat{2}\hat{3}} & =(h_{02}-h_{23})e^{-\theta},\label{eq:36a}\\
h_{\hat{0}\hat{2}}+h_{\hat{2}\hat{3}} & =(h_{02}+h_{23})e^{\theta}.\label{eq:36b}\end{align}
 \end{subequations}

The details of the calculation of the kernel are presented in Appendix~\ref{sec:B}.
We find, based on Eq.~\eqref{eq:10}, that the components of the
wave amplitude as measured by a linearly accelerated observer are
given by\begin{equation}
H_{\hat{1}\hat{1}}=h_{\hat{1}\hat{1}},\quad H_{\hat{1}\hat{2}}=h_{\hat{1}\hat{2}},\quad H_{\hat{2}\hat{2}}=h_{\hat{2}\hat{2}},\label{eq:37}\end{equation}
 \begin{subequations}\begin{align}
H_{\hat{0}\hat{0}}-H_{\hat{3}\hat{3}} & =h_{\hat{0}\hat{0}}-h_{\hat{3}\hat{3}},\label{eq:38a}\\
H_{\hat{0}\hat{0}}+H_{\hat{3}\hat{3}} & =h_{\hat{0}\hat{0}}+h_{\hat{3}\hat{3}}-\frac{4}{c}\int_{0}^{\tau}gh_{\hat{0}\hat{3}}d\tau',\label{eq:38b}\\
H_{\hat{0}\hat{3}} & =h_{\hat{0}\hat{3}}-\frac{1}{c}\int_{0}^{\tau}g(h_{\hat{0}\hat{0}}+h_{\hat{3}\hat{3}})d\tau'\label{eq:38c}\end{align}
 \end{subequations}\begin{subequations}\begin{align}
H_{\hat{0}\hat{1}}-H_{\hat{1}\hat{3}} & =h_{\hat{0}\hat{1}}-h_{\hat{1}\hat{3}}+\frac{1}{c}\int_{0}^{\tau}g(h_{\hat{0}\hat{1}}-h_{\hat{1}\hat{3}})d\tau',\label{eq:39a}\\
H_{\hat{0}\hat{1}}+H_{\hat{1}\hat{3}} & =h_{\hat{0}\hat{1}}+h_{\hat{1}\hat{3}}-\frac{1}{c}\int_{0}^{\tau}g(h_{\hat{0}\hat{1}}+h_{\hat{1}\hat{3}})d\tau',\label{eq:39b}\end{align}
 \end{subequations}\begin{subequations}\begin{align}
H_{\hat{0}\hat{2}}-H_{\hat{2}\hat{3}} & =h_{\hat{0}\hat{2}}-h_{\hat{2}\hat{3}}+\frac{1}{c}\int_{0}^{\tau}g(h_{\hat{0}\hat{2}}-h_{\hat{2}\hat{3}})d\tau',\label{eq:40a}\\
H_{\hat{0}\hat{2}}+H_{\hat{2}\hat{3}} & =h_{\hat{0}\hat{2}}+h_{\hat{2}\hat{3}}-\frac{1}{c}\int_{0}^{\tau}g(h_{\hat{0}\hat{2}}+h_{\hat{2}\hat{3}})d\tau'.\label{eq:40b}\end{align}
 \end{subequations}

An immediate consequence of these equations may be noted: For an incident
gravitational wave of definite helicity~\eqref{eq:25} propagating
along the direction of motion of the observer, $H_{\hat{i}\hat{j}}=h_{ij}$,
as follows immediately from Eqs.~\eqref{eq:33} and \eqref{eq:37}.
Thus there is no coupling of the observer's acceleration with the
helicity of the incident gravitational radiation. This is a generalization
of previous results~\cite{key-3} to nonlocal gravitation.

An important consequence of the results of this section should be
noted: The general character of the nonlocal relations for observers
that are linearly accelerated in the $z$ direction makes it possible
in this case to develop the nonlocal wave equation for linear gravitational
waves following the same steps as in~\cite{key-10} for nonlocal
Maxwell's equations. Einstein's heuristic principle of equivalence
may then be employed to argue intuitively that nonlocality should
extend to purely gravitational situations as well resulting in nonlocal
as well as nonlinear gravitational field equations.

\section{Discussion\label{sec:5}}

There is only indirect evidence at present, based on the orbital decay
of certain binary pulsars, for the existence of gravitational waves.
Assuming that gravitation involves a basic radiation field, the nonlocal
theory of accelerated observers has been extended in this paper to
include linear gravitational waves. Following the approach presented
in~\cite{key-10} for electrodynamics, it is in principle possible
to develop nonlocal field equations for linear gravitational waves
in Minkowski spacetime. This has been done in the present paper for
a rather simple class of uniformly rotating observers. Invoking Einstein's
principle of equivalence, the results of this paper may be considered
to be a step in the direction of a nonlocal classical theory of gravitation.

\appendix

\section{}

\label{sec:A}

For a general uniformly rotating observer considered in Section~\ref{sec:2},
Eq.~\eqref{eq:9} may be written out in component form as follows:\begin{align}
h_{\hat{0}\hat{0}} & =\gamma^{2}[h_{00}+\beta(-\sin\phi\; h_{01}+\cos\phi\; h_{02})+\beta^{2}(\sin^{2}\phi\; h_{11}-\sin2\phi\; h_{12}+\cos^{2}\phi\; h_{22})],\label{eq:A1}\\
h_{\hat{0}\hat{1}} & =\gamma\left[\cos\phi\; h_{01}+\sin\phi\; h_{02}+\frac{1}{2}\beta(-\sin2\phi\; h_{11}+2\cos2\phi\; h_{12}+\sin2\phi\; h_{22})\right],\label{eq:A2}\\
h_{\hat{0}\hat{2}} & =\gamma^{2}\beta\left[h_{00}+\left(\frac{1}{\beta}+\beta\right)(-\sin\phi\; h_{01}+\cos\phi\; h_{02})+\sin^{2}\phi\; h_{11}-\sin2\phi\; h_{12}+\cos^{2}\phi\; h_{22}\right],\label{eq:A3}\\
h_{\hat{0}\hat{3}} & =\gamma[h_{03}+\beta(-\sin\phi\; h_{13}+\cos\phi\; h_{23})],\label{eq:A4}\\
h_{\hat{1}\hat{1}} & =\cos^{2}\phi\; h_{11}+\sin2\phi\; h_{12}+\sin^{2}\phi\; h_{22},\label{eq:A5}\\
h_{\hat{1}\hat{2}} & =\gamma\left[\beta(\cos\phi\; h_{01}+\sin\phi\; h_{02})+\frac{1}{2}(-\sin2\phi\; h_{11}+2\cos2\phi\; h_{12}+\sin2\phi\; h_{22})\right],\label{eq:A6}\\
h_{\hat{1}\hat{3}} & =\cos\phi\; h_{13}+\sin\phi\; h_{23},\label{eq:A7}\\
h_{\hat{2}\hat{2}} & =\gamma^{2}[\beta^{2}h_{00}+2\beta(-\sin\phi\; h_{01}+\cos\phi\; h_{02})+\sin^{2}\phi\; h_{11}-\sin2\phi\; h_{12}+\cos^{2}\phi\; h_{22}],\label{eq:A8}\\
h_{\hat{2}\hat{3}} & =\gamma(\beta h_{03}-\sin\phi\; h_{13}+\cos\phi\; h_{23}),\label{eq:A9}\\
h_{\hat{3}\hat{3}} & =h_{33}.\label{eq:A10}\end{align}
 These equations are equally valid for uniformly rotating observers
at any fixed value of the vertical coordinate $z$.

For the incident radiation field~\eqref{eq:25} as measured by rotating
observers at $z$, Eqs.~\eqref{eq:A1}-\eqref{eq:A10} imply that
\begin{equation}
(h_{\hat{\alpha}\hat{\beta}})=A\left[\begin{array}{cccc}
-\beta^{2}\gamma^{2} & \pm i\beta\gamma & -\beta\gamma^{2} & 0\\
\pm i\beta\gamma & 1 & \pm i\gamma & 0\\
-\beta\gamma^{2} & \pm i\gamma & -\gamma^{2} & 0\\
0 & 0 & 0 & 0\end{array}\right]e^{\pm2i\phi}e^{i\omega(-t+z/c)},\label{eq:A11}\end{equation}
 where the temporal dependence is of the form $\exp(-i\omega'\tau)$
with $\omega'$ given by Eq.~\eqref{eq:1}. Equation~\eqref{eq:A11}
should be compared and contrasted with the measured components of
the Riemann tensor in this case given by Eqs.~(2.10) and (2.11) of~\cite{key-1}.
Moreover, we note that for the special rotating observer with $\beta=0$
and $\gamma=1$, Eq.~\eqref{eq:A11} simply reduces to Eq.~\eqref{eq:27}.

\section{}

\label{sec:B}

The purpose of this appendix is to compute $\Lambda^{-1}$ and the
kernel $k$ for the linearly accelerated observers of Section~\ref{sec:4}.
Inspection of $\Lambda$ reveals that the entries in its fifth and
sixth rows and columns vanish except for the diagonal elements that
both equal unity. Let the $8\times8$ matrix $\tilde{\Lambda}$ be
the reduced form of $\Lambda$ in which the fifth and sixth rows and
columns have been ignored; then,\begin{equation}
\tilde{\Lambda}(\theta)=\left[\begin{array}{cc}
M & N\\
P & Q\end{array}\right],\label{eq:B1}\end{equation}
 where \begin{align}
M & =\left[\begin{array}{cccc}
C^{2} & 0 & 0 & 2CS\\
0 & C & 0 & 0\\
0 & 0 & C & 0\\
CS & 0 & 0 & C^{2}+S^{2}\end{array}\right], & \quad N & =S\left[\begin{array}{cccc}
0 & 0 & 0 & S\\
1 & 0 & 0 & 0\\
0 & 0 & 1 & 0\\
0 & 0 & 0 & C\end{array}\right],\label{eq:B2}\\
P & =S\left[\begin{array}{cccc}
0 & 1 & 0 & 0\\
0 & 0 & 0 & 0\\
0 & 0 & 1 & 0\\
S & 0 & 0 & 2C\end{array}\right], & \quad Q & =\diag(C,1,C,C^{2}).\label{eq:B3}\end{align}
 Here, as before, $C=\cosh\theta$ and $S=\sinh\theta$. We note that
$\det M=\det Q=C^{4}$ and $\det N=\det P=0$. It can be shown that\begin{equation}
\tilde{\Lambda}^{-1}(\theta)=\left[\begin{array}{cc}
U & V\\
X & Y\end{array}\right],\label{eq:B4}\end{equation}
 where\begin{align}
U & =(M-NQ^{-1}P)^{-1}, & \quad V & =-M^{-1}NY,\label{eq:B5}\\
X & =-Q^{-1}PU, & \quad Y & =(Q-PM^{-1}N)^{-1}.\label{eq:B6}\end{align}
 From\begin{equation}
M^{-1}=\frac{1}{C^{2}}\left[\begin{array}{cccc}
C^{2}+S^{2} & 0 & 0 & -2CS\\
0 & C & 0 & 0\\
0 & 0 & C & 0\\
-CS & 0 & 0 & C^{2}\end{array}\right]\label{eq:B7}\end{equation}
 and the explicit evaluation of the matrices in Eqs.~\eqref{eq:B5}
and \eqref{eq:B6}, one finds the simple relation\begin{equation}
\tilde{\Lambda}^{-1}(\theta)=\tilde{\Lambda}(-\theta).\label{eq:B8}\end{equation}
 The same relation holds for $\Lambda$, i.e. $\Lambda^{-1}$ has
the same form as $\Lambda$ but with $(C,S)\mapsto(C,-S)$. Moreover,
the nonzero elements of $k$ in this case turn out to be\begin{align}
k_{1,4} & =k_{10,4}=-\frac{2}{c}g(\tau),\label{eq:B9}\\
k_{2,7} & =k_{3,9}=k_{4,1}=k_{4,10}=k_{7,2}=k_{9,3}=-\frac{1}{c}g(\tau).\label{eq:B10}\end{align}

\end{document}